\documentclass[seceq,supplement]{ptptex}

\usepackage{amsmath}
\usepackage{graphicx}
\usepackage{amssymb}

\title{
Light-cone Gauge String Field Theory\\ and Dimensional Regularization%
}

\author{
Nobuyuki \textsc{Ishibashi}$^{1,}$\footnote{%
  e-mail: ishibash@het.ph.tsukuba.ac.jp}
and
Koichi \textsc{Murakami}$^{2,}$\footnote{%
  e-mail: koichimurakami71@gmail.com}
}

\inst{
${}^{1}$Institute of Physics, University of Tsukuba,
Tsukuba, Ibaraki 305-8571, Japan\\
${}^{2}$Okayama Institute for Quantum Physics, 
Kyoyama 1-9-1, Kita-ku, \\
Okayama 700-0015, Japan
}

\abst{
We review our recent proposals to dimensionally regularize the light-cone
gauge string field theory.
}

\begin{document}

\maketitle

\section{Introduction}

One of the biggest problems in string field theory is how to treat
the contact term divergences\ \cite{Greensite:1986gv,Greensite:1987hm,Greensite:1987sm,Green:1987qu,Wendt:1987zh}.
There are several proposals to do so for Witten's open string field
theory\ \cite{Arefeva1990a,Preitschopf1990,Berkovits1995}. We have
shown that the dimensional regularization can be employed to deal
with it\ \cite{Baba:2009kr,Baba:2009ns,Baba:2009fi,Baba:2009zm,Ishibashi:2010nq}
in the case of the light-cone gauge string field theory. 
It is possible to formulate the light-cone gauge string field theory 
in noncritical spacetime dimension $d$. We can define scattering amplitudes 
as analytic functions of $d$ and obtain those in the critical dimensions 
by taking the limit $d\to 10$. 
In this note,
we would like to outline the procedure, focusing on points which 
were not discussed explicitly in the original references.

\section{Light-cone gauge SFT}

Let us start by recapitulating how to define the light-cone gauge 
string field theory in critical dimensions. 

\subsection{Light-cone gauge SFT action}

Light-cone gauge string field theory for closed bosonic strings can
be described by the action\ \cite{Kaku:1974zz,Kaku:1974xu} \begin{eqnarray}
S & = & \int\left[\frac{1}{2}\Phi\cdot K\Phi+\frac{g}{6}\Phi\cdot\left(\Phi\ast\Phi\right)\right]\ .\label{eq:action}\end{eqnarray}
Here the string field $\Phi$ is an element of the Hilbert space of
the transverse variables $X^{i}$ which satisfies the level matching
condition:\begin{eqnarray*}
\left|\Phi\left(t,\alpha\right)\right\rangle  & \in & \mathcal{H}_{X^{i}}\ \left(i=1,\cdots,d-2=24\right)\ ,\\
 &  & \left(L_{0}-\tilde{L}_{0}\right)\left|\Phi\left(t,\alpha\right)\right\rangle =0\ .\end{eqnarray*}
It is a function of $t\equiv x^{+},\alpha=2p^{+}$. The inner product
of the string fields is defined as \begin{eqnarray*}
\int\Phi_{1}\cdot\Phi_{2} & \equiv & \int dt\frac{\alpha d\alpha}{4\pi}\left\langle \Phi_{1}\left(t,-\alpha\right)|\Phi_{2}\left(t,\alpha\right)\right\rangle \ .\end{eqnarray*}
The kinetic operator $K$ is given as \begin{eqnarray*}
K & \equiv & i\partial_{t}-\frac{L_{0}+\tilde{L}_{0}-\frac{d-2}{12}}{\alpha}\ .\end{eqnarray*}

\begin{figure}
\begin{centering}
\includegraphics[scale=0.5]{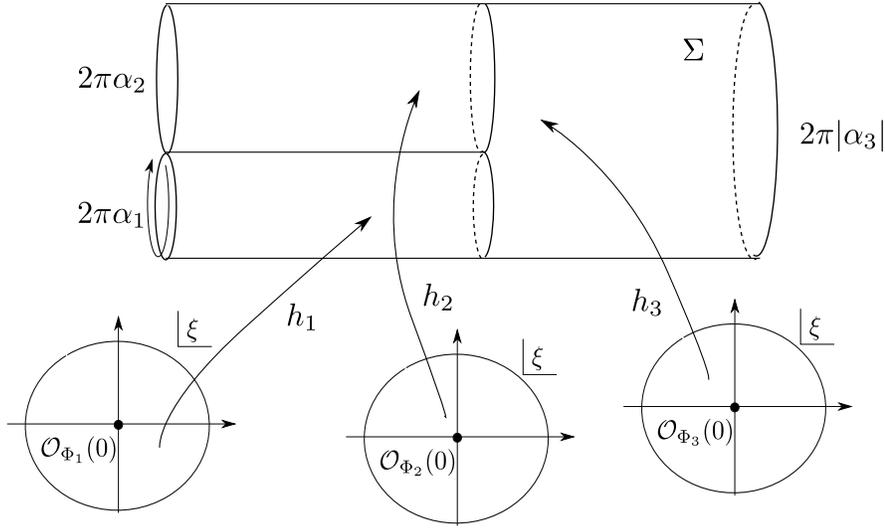}
\par\end{centering}

\caption{Three string vertex for $\alpha_{1},\alpha_{2}>0,\,\alpha_{3}<0$\label{fig:three-string-vertex}}

\end{figure}

The action includes only a three string interaction term. The interaction
term can be defined by using the state-operator correspondence. To
a string field $\left|\Phi\right\rangle $, there corresponds a local
operator $\mathcal{O}_{\Phi}(z)$ such that \begin{eqnarray*}
\left|\Phi\right\rangle  & = & \mathcal{O}_{\Phi}(0)\left|0\right\rangle \ ,\end{eqnarray*}
where $\left|0\right\rangle $ is the SL($2,\mathbb{C}$) invariant
vacuum. Then the integral for three string fields can be given as
\begin{eqnarray}
\int\Phi_{1}\cdot\left(\Phi_{2}\ast\Phi_{3}\right) & = & \int dt\prod_{r=1}^{3}\left(\frac{\alpha_{r}d\alpha_{r}}{4\pi}\right)\delta\left(\sum_{r=1}^{3}\alpha_{r}\right)\nonumber \\
 &  & \hphantom{\int dt}\times\left\langle h_{1}\circ\mathcal{O}_{\Phi_{1}\left(t,\alpha_{1}\right)}h_{2}\circ\mathcal{O}_{\Phi_{2}\left(t,\alpha_{2}\right)}h_{3}\circ\mathcal{O}_{\Phi_{3}\left(t,\alpha_{3}\right)}\right\rangle _{\Sigma}\ ,\label{eq:three-string}\end{eqnarray}
where $h_{r}\,\left(r=1,2,3\right)$ are the maps which are depicted
in Fig.~\ref{fig:three-string-vertex}. 

\begin{figure}[h]
\begin{centering}
\includegraphics[scale=0.5]{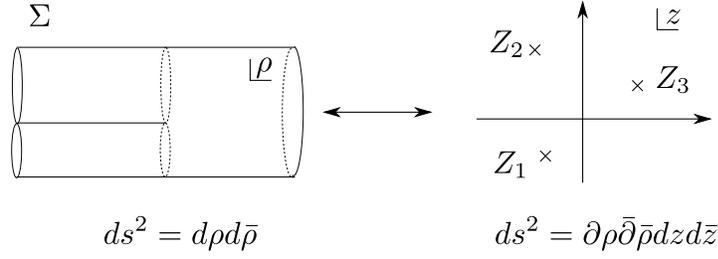}
\par\end{centering}

\caption{Mandelstam mapping\label{fig:Mandelstam}}

\end{figure}

The worldsheet theory for the light-cone gauge SFT possesses nonvanishing
Virasoro central charge even in the critical case. This fact makes
the calculation of the correlation function on the right hand side
of eq.(\ref{eq:three-string}) a little bit complicated. It
can be evaluated by using the Mandelstam mapping\ \cite{Mandelstam:1973jk}
\begin{eqnarray*}
\rho\left(z\right) & = & \sum_{r=1}^{3}\alpha_{r}\ln\left(z-Z_{r}\right)\ ,\end{eqnarray*}
which maps the complex plane to $\Sigma$ as is described in 
Fig.~\ref{fig:Mandelstam}.
Using this, we can rewrite
the correlation functions in terms of those on the complex plane.
Because of the conformal anomaly, we obtain\begin{eqnarray*}
 &  & \left\langle h_{1}\circ\mathcal{O}_{\Phi_{1}}h_{2}\circ\mathcal{O}_{\Phi_{2}}h_{3}\circ\mathcal{O}_{\Phi_{3}}\right\rangle _{\Sigma}\\
 &  & \quad=\left\langle \left(\rho^{-1}h_{1}\right)\circ\mathcal{O}_{\Phi_{1}}\left(\rho^{-1}h_{2}\right)\circ\mathcal{O}_{\Phi_{2}}\left(\rho^{-1}h_{3}\right)\circ\mathcal{O}_{\Phi_{3}}\right\rangle _{\mathbb{C}}e^{-\Gamma\left[\ln\left(\partial\rho\bar{\partial}\bar{\rho}\right)\right]}\ ,\end{eqnarray*}
where \begin{eqnarray}
\Gamma\left[\phi\right] & = & -\frac{1}{\pi}\int d^{2}z\partial\phi\bar{\partial}\phi  \label{eq:Liouville}\end{eqnarray}
is the Liouville action. Using this form, the right hand side of eq.(\ref{eq:three-string})
is given as \begin{eqnarray*}
 &  & \int\Phi_{1}\cdot\left(\Phi_{2}\ast\Phi_{3}\right)\\
 &  & \quad=\int dt\prod_{r=1}^{3}\left(\frac{\alpha_{r}d\alpha_{r}}{4\pi}\right)\delta\left(\sum_{r=1}^{3}\alpha_{r}\right)\\
 &  & \hphantom{\quad=\int\prod_{r=1}^{3}}\times\left\langle \left(\rho^{-1}h_{1}\right)\circ\mathcal{O}_{\Phi_{1}}\left(\rho^{-1}h_{2}\right)\circ\mathcal{O}_{\Phi_{2}}\left(\rho^{-1}h_{3}\right)\circ\mathcal{O}_{\Phi_{3}}\right\rangle _{\mathbb{C}}\\
 &  & \hphantom{\quad=\int\prod_{r=1}^{3}}\times e^{-\Gamma\left[\ln\left(\partial\rho\bar{\partial}\bar{\rho}\right)\right]}\ .\end{eqnarray*}
The Liouville action $\Gamma\left[\ln\left(\partial\rho\bar{\partial}\bar{\rho}\right)\right]$
for the Mandelstam mapping can be calculated by using the method explained
later. In this case, we obtain\begin{eqnarray}
e^{-\Gamma\left[\ln\left(\partial\rho\bar{\partial}\bar{\rho}\right)\right]} & = & \frac{\exp\left(-2\sum_{r}\frac{\hat{\tau}_{0}}{\alpha_{r}}\right)}{\alpha_{1}\alpha_{2}\alpha_{3}}\ ,\label{eq:Gamma3}\\
\hat{\tau}_{0} & \equiv & \sum_{r=1}^{3}\alpha_{r}\ln\left|\alpha_{r}\right|\ .\nonumber \end{eqnarray}
Therefore the three string vertex in the light-cone gauge consists
of LPP part and the part which comes from the anomaly.

\subsection{Amplitudes}

\begin{figure}[h]
\begin{centering}
\includegraphics[scale=0.5]{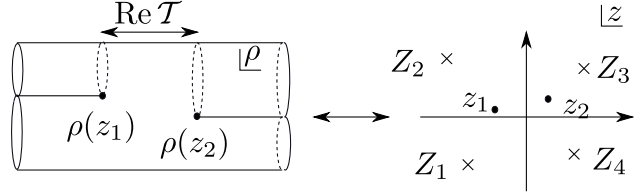}
\par\end{centering}

\caption{Tree level four string amplitude\label{fig:tree-level-four}}

\end{figure}

Amplitudes in the light-cone gauge SFT can be evaluated by using the
propagator and the vertex derived form the action (\ref{eq:action}).
The tree level four string amplitudes correspond to the Feynman diagram
depicted in Fig.~\ref{fig:tree-level-four}. We can evaluate tree
level $N$ string amplitudes by mapping the worldsheet to the complex
plane by the Mandelstam mapping\begin{eqnarray}
\rho\left(z\right) & = & \sum_{r=1}^{N}\alpha_{r}\ln\left(z-Z_{r}\right)\ .\label{eq:rhoN}\end{eqnarray}
As in the three string vertex, the amplitudes can be written in terms
of the correlation functions on the complex plane as\begin{eqnarray}
\mathcal{A}_{N} & = & \sum_{\mathrm{channels}}\int\prod_{\mathcal{I}}d^{2}\mathcal{T}_{\mathcal{I}}\left\langle \prod_{r}V_{r}^{\mathrm{LC}}\right\rangle _{\mathbb{C}}e^{-\Gamma\left[\ln\partial\rho\bar{\partial}\bar{\rho}\right]}\ .\label{eq:AN}\end{eqnarray}
Here $V_{r}^{\mathrm{LC}}$ is the vertex operator corresponding to
the $r$-th external line, whose explicit form is given in 
Ref.~\citen{Baba:2009ns}. $e^{-\Gamma\left[\ln\partial\rho\bar{\partial}\bar{\rho}\right]}$
is the factor coming from the conformal anomaly.

\subsection{Evaluation of $-\Gamma$}

\begin{figure}[h]
\begin{centering}
\includegraphics{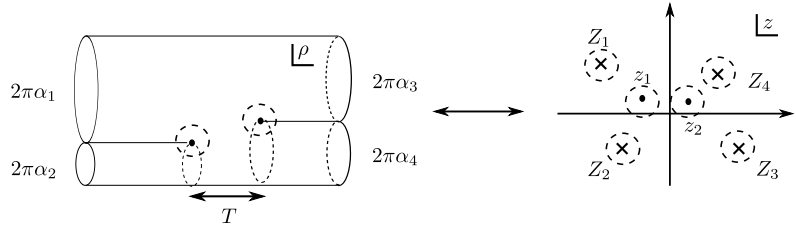}
\par\end{centering}

\caption{Mandelstam's regularization\label{fig:regularization}}

\end{figure}

In principle, $-\Gamma$ can be calculated by substituting (\ref{eq:rhoN})
into (\ref{eq:Liouville}). However, $\partial\phi$ diverges at $z=Z_{r},\, z_{I}$,
where $z_{I}\,\left(I=1,\cdots,N-2\right)$ are the coordinates of
the interaction points on the complex plane. In order to get a meaningful
result, we need to regularize the divergences. A method to perform
such calculations were given by Mandelstam\  \cite{Mandelstam:1985ww}.
Here we briefly review his method. An alternative way to evaluate
$-\Gamma$ is given in Appendix C of Ref.~\citen{Baba:2009ns}. 

In order to regularize the divergences, we consider the complex plane
with small discs around $z=Z_{r},z_{I},\infty$ with radii $\epsilon_{r},\epsilon_{I},\epsilon_{\infty}$
excised as shown in Fig.~\ref{fig:regularization}
and evaluate the Liouville action on this surface with boundaries.
$-\Gamma$ diverges in the limit $\epsilon_{r},\epsilon_{I},\epsilon_{\infty}\to0$. 
We would like to calculate $-\Gamma$ keeping the reparametrization invariance and add reparametrization invariant and local counterterms to make it finite. 

$-\Gamma$ can be given as a sum of the contributions from the boundaries.
The Liouville action for $\phi$ on the surface can be expressed as
a sum over surface terms. With discs excised, the Liouville action
for the flat metric is given as a function of $\epsilon_{r},\epsilon_{I},\epsilon_{\infty}$.
Since these cutoff parameters are not reparametrization invariant
as we will see, we should add the flat metric contribution to keep
the reparametrization invariance.

For example, let us consider the disc around $z=Z_{r}$. Excising
a disc of radius $\epsilon_{r}\ll1$ around $z=Z_{r}$ corresponds
to making the external cylinder of the $r$-th external line to be
of the length $\alpha_{r}T_{r}\gg1$. $T_{r}$ is invariant under
the reparametrization because $\rho$ is transformed as a scalar.
The relation between $\epsilon_{r}$ and $T_{r}$ is given as \begin{eqnarray}
\epsilon_{r} & \sim & e^{-T_{r}+\mathrm{Re}\bar{N}_{00}^{rr}}\ ,\nonumber \\
\bar{N}_{00}^{rr} & \equiv & \frac{\rho\left(z_{I^{\left(r\right)}}\right)}{\alpha_{r}}-\sum_{s\ne r}\frac{\alpha_{s}}{\alpha_{r}}\ln\left(Z_{r}-Z_{s}\right)\ .\label{eq:Nbarrr}\end{eqnarray}
Here $z_{I^{\left(r\right)}}$ denotes the interaction point where
the $r$-th external line interacts. The contribution to the Liouville
action for $\phi$ is given as\[
-2\ln\left|\alpha_{r}\right|+2\ln\epsilon_{r}\ ,\]
and that to the Liouville action for the flat metric is\[
-4\ln\epsilon_{r}\ .\]
Therefore the contribution of this boundary to $e^{-\Gamma}$ is given as
\[
\frac{e^{-2\mathrm{Re}\bar{N}_{00}^{rr}}}{\left|\alpha_{r}\right|^{2}}e^{2T_{r}}\ .\]

Other boundaries can be treated in the same way. The disc of radius
$\epsilon_{I}\ll1$ around $z=z_{I}$ on the complex plane corresponds
to a disc of radius $r_{I}$ on the $\rho$ plane and \begin{eqnarray*}
r_{I} & \sim & \frac{1}{2}\left|\partial^{2}\rho\left(z_{I}\right)\right|\epsilon_{I}^{2}\ .\end{eqnarray*}
The contribution to $e^{-\Gamma}$ is given as \[
\frac{1}{\left|\partial^{2}\rho\left(z_{I}\right)\right|\left(2r_{I}\right)^{5}}\ .\]
The disc of radius $\epsilon_{\infty}\ll 1$ around $z=\infty$
on the complex plane corresponds to a disc of radius $r_{\infty}$
on the $\rho$ plane and \begin{eqnarray*}
r_{\infty} & \sim & \left|\sum_{r}\alpha_{r}Z_{r}\right|\epsilon_{\infty}\ .\end{eqnarray*}
The contribution to $e^{-\Gamma}$ is given as \[
\left|\sum_{r}\alpha_{r}Z_{r}\right|^{4}r_{\infty}^{-4}\ .\]
$r_{I},r_{\infty}$ are the reparametrization invariant cutoff parameters.
Putting
all these together, we get \begin{eqnarray*}
e^{-\Gamma} & \sim & \frac{\left|\sum_{s}\alpha_{s}Z_{s}\right|^{4}e^{-2\sum_{r}\mathrm{Re}\bar{N}_{00}^{rr}}}{\prod_{r}\left|\alpha_{r}\right|^{2}\prod_{I}\left|\partial^{2}\rho\left(z_{I}\right)\right|}\cdot r_{\infty}^{-4}\prod_{r}e^{2T_{r}}\prod_{I}\left(2r_{I}\right)^{-5}\ .\end{eqnarray*}
Divergent factor $r_{\infty}^{-4}\prod_{r}e^{2T_{r}}\prod_{I}\left(2r_{I}\right)^{-5}$
can be absorbed into the normalizations of the coupling constant $g$
and the wave functions of the external lines. Thus we get the following
form of $e^{-\Gamma}$:
\begin{eqnarray}
e^{-\Gamma\left[\ln\left(\partial\rho\bar{\partial}\bar{\rho}\right)\right]} & = & \frac{\left|\sum_{s}\alpha_{s}Z_{s}\right|^{4}e^{-2\sum_{r}\mathrm{Re}\bar{N}_{00}^{rr}}}{\prod_{r}\left|\alpha_{r}\right|^{2}\prod_{I}\left|\partial^{2}\rho\left(z_{I}\right)\right|}\ .\label{eq:e-Gamma}\end{eqnarray}
For $N=3$, the right hand side coincides with eq.(\ref{eq:Gamma3}). 
The normalization for general $N$
can be fixed by examining the factorization properties\ \cite{Baba:2009kr}. 

The right hand side of eq.(\ref{eq:AN}) can be turned into an integral
over the moduli space. With the form of $-\Gamma$ given in (\ref{eq:e-Gamma}),
one can show that the amplitudes given in eq.(\ref{eq:AN}) 
coincide with the usual first-quantized results \cite{Baba:2009ns}.

\section{Light-cone gauge SFT in noncritical dimensions}

There is no problem in defining the light-cone gauge SFT for $d\ne26$.
We write down the action \begin{eqnarray*}
S & = & \int\left[\frac{1}{2}\Phi\cdot K\Phi+\frac{g}{6}\Phi\cdot\left(\Phi\ast\Phi\right)\right]\ ,\end{eqnarray*}
where this time \begin{eqnarray*}
K & \equiv & i\partial_{t}-\frac{L_{0}+\tilde{L}_{0}-\frac{d-2}{12}}{\alpha}\ ,\end{eqnarray*}
and \begin{eqnarray*}
 &  & \int\Phi_{1}\cdot\left(\Phi_{2}\ast\Phi_{3}\right)\\
 &  & \quad=\int dt\prod_{r=1}^{3}\left(\frac{\alpha_{r}d\alpha_{r}}{4\pi}\right)\delta\left(\sum_{r=1}^{3}\alpha_{r}\right)\\
 &  & \hphantom{\quad=\int\prod_{r=1}^{3}}\times\left\langle \left(\rho^{-1}h_{1}\right)\circ\mathcal{O}_{\Phi_{1}}\left(\rho^{-1}h_{2}\right)\circ\mathcal{O}_{\Phi_{2}}\left(\rho^{-1}h_{3}\right)\circ\mathcal{O}_{\Phi_{3}}\right\rangle _{\mathbb{C}}\\
 &  & \hphantom{\quad=\int\prod_{r=1}^{3}}\times e^{-\frac{d-2}{24}\Gamma\left[\ln\left(\partial\rho\bar{\partial}\bar{\rho}\right)\right]}\ .\end{eqnarray*}
The amplitudes are calculated in the same way as those in the critical
case:\begin{eqnarray}
\mathcal{A}_{N} & = & \sum_{\mathrm{channels}}\int\prod_{\mathcal{I}}d^{2}\mathcal{T}_{\mathcal{I}}\left\langle \prod_{r}V_{r}^{\mathrm{LC}}\right\rangle _{\mathbb{C}}e^{-\frac{d-2}{24}\Gamma\left[\ln\left(\partial\rho\bar{\partial}\bar{\rho}\right)\right]}\ .\label{eq:ANnon}\end{eqnarray}

We would like to turn the form of the amplitude in eq.(\ref{eq:ANnon})
into an integral over the moduli space. The moduli parameters can
be taken to be the positions of the vertex operators on the complex
plane. In doing so, there are two points to be checked. 

One thing is the SL($2,\mathbb{C}$) invariance. The integrand on
the right hand side is given in terms of the quantities defined on
the complex plane, but it was originally defined on the $\rho$ plane.
Therefore we expect that  we eventually obtain an SL($2,\mathbb{C}$)
invariant expression. Indeed it is easy to check the invariance.\footnote{The
one-loop amplitudes can be shown to be invariant under the modular
transformation \cite{Murakami}.}

\begin{figure}
\begin{centering}
\includegraphics[scale=0.5]{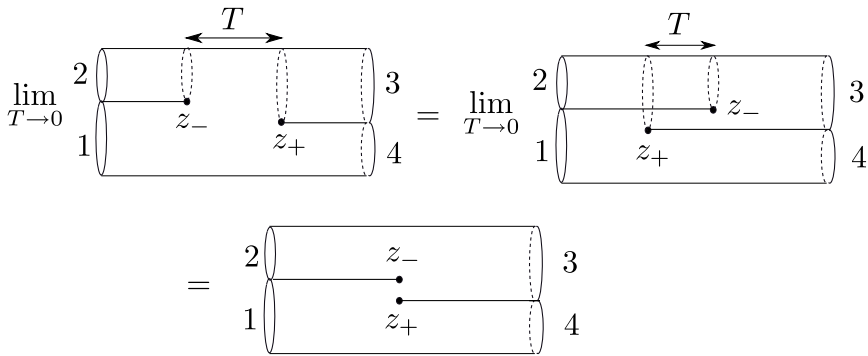}
\par\end{centering}

\caption{Contributions from various channels\label{fig:channels}}

\end{figure}

Another thing to be checked is if the contributions from various channels
are smoothly connected. In string field theory, the amplitudes are
given as a sum over contributions from various channels. Each channel
corresponds to a region in the moduli space. We need to check if the
integrand is smoothly connected at the boundaries of these regions
as indicated in Fig.~\ref{fig:channels} so that the $\mathcal{A}_{N}$
is given as an integral of a smooth function over the moduli space.
If the integrand is discontinuous at the boundaries, we have troubles in proving
various symmetries of the amplitudes. It is easy to check that there are no discontinuities 
and we obtain an integral of a smooth function over the moduli space.

Thus the amplitudes for the noncritical string field theory can be
constructed without any trouble. Therefore it seems that there is
nothing wrong in considering the string field theory in noncritical
dimensions.

\section{Contact term divergences and dimensional regularization}

The noncritical string field theory can be used to regularize the
contact term divergences. It is possible to define light-cone gauge
superstring field theory in noncritical dimensions as in the bosonic
case. Taking $d\ne10$ naively, we obtain a theory with only spacetime
bosons which cannot be used to regularize the amplitudes with spacetime
fermions. In order to deal with fermions, we need to modify the worldsheet
theory. Details of such a treatment will be given elsewhere. 

The amplitudes can be given as \begin{equation}
\mathcal{A}_{N}=\sum_{\mathrm{channels}}\int\prod_{\mathcal{I}}d^{2}\mathcal{T}_{\mathcal{I}}\left\langle \prod_{I}\left|\left(\partial^{2}\rho\right)^{-\frac{3}{4}}T_{F}\left(z_{I}\right)\right|^{2}\prod_{r}V_{r}^{\mathrm{LC}}\right\rangle _{\mathbb{C}}e^{-\frac{d-2}{16}\Gamma\left[\ln\left(\partial\rho\bar{\partial}\bar{\rho}\right)\right]}\ .\label{eq:superAN}\end{equation}
Compared with the bosonic case (\ref{eq:ANnon}), the big difference
is the existence of the transverse supercurrent $T_{F}$ at the interaction
points. When some interaction points come close to each other, the
correlation function diverges for $d=10$ and we cannot define $\mathcal{A}_{N}$
even for tree amplitudes. 
However, it can be shown that for $\left|z_{I}-z_{J}\right|\sim0$,\[
e^{-\frac{d-2}{16}\Gamma\left[\ln\left(\partial\rho\bar{\partial}\bar{\rho}\right)\right]}\sim\left|z_{I}-z_{J}\right|^{-\frac{d-2}{8}}\ ,\]
and taking $d$ largely negative we can make $\mathcal{A}_{N}$ well-defined.
We can define $\mathcal{A}_{N}$ as an analytic function of $d$
and take the limit $d\to10$. If the limit is finite, 
a definition of $\mathcal{A}_{N}$ for $d=10$ can be obtained. 
If the limit is divergent,
we should add counterterms to make it finite.

\section{Conformal gauge formulation\label{sec:Conformal-gauge-formulation}}

One can show that the tree amplitudes defined by the dimensional regularization
are finite in the limit $d\to10$ and reproduce the results of the
first-quantized formalism. This fact can be shown by constructing
the conformal gauge formulation of the noncritical strings. 

In the conformal gauge, the noncritical light-cone gauge strings should
correspond to a worldsheet theory in a Lorentz noninvariant background.
The conformal gauge formulation can be constructed by the following
reasonings. The light-cone gauge worldsheet theory can be described
by the path integral\begin{equation}
\int\left[dX^{i}\right]e^{-S^{\mathrm{LC}}}\ ,\label{eq:LCpath}\end{equation}
where \[
S^{\mathrm{LC}}=\frac{1}{2\pi}\int d^{2}z\partial X^{i}\bar{\partial}X^{i} \]
is the worldsheet action for the transverse variables $X^{i}\,\left(i=1,\cdots,d-2\right)$.
This action can be considered as the gauge fixed version of the standard
Nambu-Goto action $S^{\mathrm{NG}}$ and the path integral should
correspond to \begin{equation}
\int\left[dX^{\mu}\right]e^{-S^{\mathrm{NG}}}\ .\label{eq:NGpath}\end{equation}
For $d\ne26$, we need to specify the worldsheet metric which should
be used to define the path integral measure $\left[dX^{\mu}\right]$.
In the light-cone gauge, the natural metric on the worldsheet is \begin{eqnarray*}
ds^{2} & = & d\rho d\bar{\rho}\\
 & \sim & \partial X^{+}\bar{\partial}X^{+}dzd\bar{z}\ .\end{eqnarray*}
Therefore the path integral (\ref{eq:NGpath}) should be written as
\begin{equation}
\int\left[dX^{\mu}\right]_{\partial X^{+}\bar{\partial}X^{+}}e^{-S^{\mathrm{NG}}}\ ,\label{eq:NGpath2}\end{equation}
where we have indicated the metric $\partial X^{+}\bar{\partial}X^{+}$
to be used to define the measure. In the conformal gauge, eq.(\ref{eq:NGpath2})
corresponds to the action\[
\frac{1}{2\pi}\int d^{2}z\partial X^{\mu}\bar{\partial}X_{\mu}+\frac{d-26}{24}\Gamma\left[\partial X^{+}\bar{\partial}X^{+}\right]\ .\]
For $d\ne26$, this worldsheet theory is not Lorentz invariant. 

The longitudinal part of the worldsheet theory defines a nontrivial
CFT. One can derive the energy-momentum tensor as \begin{eqnarray}
T\left(z\right) & \equiv & \partial X^{+}\partial X^{-}-\frac{d-26}{12}\left\{ X^{+},z\right\} \ ,\label{eq:emtensor}\end{eqnarray}
where \begin{eqnarray*}
\left\{ X^{+},z\right\}  & \equiv & \frac{\partial^{3}X^{+}}{\partial X^{+}}-\frac{3}{2}\left(\frac{\partial^{2}X^{+}}{\partial X^{+}}\right)^{2} \end{eqnarray*}
is the Schwarzian derivative. Such a conformal field theory can be
analyzed by using the path integral formalism \cite{Baba:2009ns}.
One can show that the energy-momentum tensor (\ref{eq:emtensor})
satisfies the Virasoro algebra with $c=28-d$. Therefore, with the
transverse part and the reparametrization ghosts, the worldsheet theory
is with nilpotent BRST charge. 

We can construct the BRST invariant conformal gauge formulation in
a similar way for superstrings. When all the external lines are
bosonic, one can rewrite the light-cone gauge amplitudes (\ref{eq:superAN})
into a BRST invariant form in the conformal gauge formulation. Using
such a form, it is possible to show that the dimensionally regularized
tree amplitudes has a finite $d\to10$ limit and the results coincide
with those of the first-quantized formalism \cite{Murakami,Ishibashi:2010nq}.

\section{Summary}

We have shown that the dimensional regularization is a useful tool
to deal with the contact term divergences of light-cone gauge string
field theory. Our method would be applicable to the gauge invariant
SFT's \cite{Hata:1986jd,Hata:1986kj} with joining-splitting type
interactions. Moreover the conformal gauge formulation in section
\ref{sec:Conformal-gauge-formulation} may shed light on the construction
of covariant SFT of this type \cite{Hata:1987qx,Kugo:1992md}.

\section*{Acknowledgements}
We would like to thank the organizers of the conference for a wonderful
meeting where many exciting developments are presented and discussed.
This work was supported in part by Grant-in-Aid for Scientific Research~(C)
(20540247) from the Ministry of Education, Culture, Sports, Science
and Technology (MEXT) of Japan.

%

\end{document}